# Hybrid Graphene/Silicon Schottky photodiode with intrinsic gating effect

Antonio Di Bartolomeo[1,2,*], Giuseppe Luongo[1], Filippo Giubileo[2], Nicola Funicello[1], Gang Niu[3], Thomas Schroeder[4,5], Marco Lisker[4], and Grzegorz Lupina[4]

[1]Physics Department "E. R. Caianiaello", Università di Salerno, via Giovanni Paolo II, Fisciano, 84084, Italy.

[2]CNR-SPIN Salerno, via Giovanni Paolo II, Fisciano, 84084, Italy

[3]Electronic Materials Research Laboratory, Key Laboratory of the Ministry of Education & International Center for Dielectric Research, Xi'an Jiaotong University, Xi'an 710049, China

[4]IHP Microelectronics, Im Technologiepark 25, 15236 Frankfurt (Oder), Germany

[5]Brandenburg University of Technology. Institute of Physics, Konrad Zuse Str. 1, 03046 Cottbus, Germany

[*] Corresponding author. E-mail: adibartolomeo@unisa.it

**Abstract**

We propose a hybrid device consisting of a graphene/silicon (Gr/Si) Schottky diode in parallel with a Gr/SiO$_2$/Si capacitor for high-performance photodetection. The device, fabricated by transfer of commercial graphene on low-doped n-type Si substrate, achieves a photoresponse as high as $3\ \text{AW}^{-1}$ and a normalized detectivity higher than $3.5 \times 10^{12}\ \text{cmHz}^{1/2}\text{W}^{-1}$ in the visible range. The device exhibits a photocurrent exceeding the forward current, because photo-generated minority carriers, accumulated at Si/SiO$_2$ interface of the Gr/SiO$_2$/Si capacitor, diffuse to the Gr/Si junction. We show that the same mechanism, when due to thermally generated carriers, although usually neglected or disregarded, causes the increased leakage often measured in Gr/Si heterojunctions. At room temperature, we measure a zero-bias Schottky barrier height of 0.52 eV, as well as an effective Richardson constant $A^{**}=4 \times 10^{-5}\ \text{Acm}^{-2}\text{K}^{-2}$ and an ideality factor $n \approx 3.6$, explained by a thin (< 1nm) oxide layer at the Gr/Si interface.

# 1. Introduction

The scientific and engineering research on the heterojunction formed by graphene with 2D or 3D semiconductors (Gr/S junction) has recently entered an hectic phase [1-6], matching or overcoming the activity on graphene field-effect transistors [7-11]. The interest in the Gr/S junction originates from unique transport properties that enable the study of fundamental physics as well as the fabrication of performant rectifiers [12,13], photodetectors [14-16], solar cells or chemical sensors [17-21]. Moreover, the Gr/S junction is a basic element of novel electronic devices for the integration of graphene into the existing semiconductor technology [22,23].

Despite the flurry of experimental and theoretical studies, the physics underlying the Gr/S junction is still incomplete. Often, the Gr/S system involves the formation of a Schottky barrier and exhibits a rectifying behaviour, but the details of the current-voltage (I-V) characteristic are strongly dependent on the quality of the interface, on the used materials and more in general on the fabrication method. While the forward current is usually described by thermionic emission theory, an unexpectedly high and bias-dependent reverse leakage current is often measured [1,24]. The anomaly of the reverse saturation current has been explained by the modulation of the Schottky barrier height due to the low density of states (DOS) of graphene [16,25]. Other inconsistencies with the standard Schottky diode model, as the lower value of the effective Richardson constant or its dependence on temperature, as well as the behavior of the ideality factor, still remain open questions [4].

In this paper, we fabricate and characterize a planar Gr/Si junction on low-doped n-type substrate. The device, which is realized by transferring a monolayer graphene strip over a narrow slit etched in the dielectric covering the Si chip (Figure 1), purposely includes a parallel wide-area metal-oxide-semiconductor (MOS) capacitor, which has profound effects on the optoelectronic behaviour of the Gr/Si junction. The MOS capacitor, which in the present application has the metal replaced by graphene and the oxide consisting of a stack of $SiO_2$-$Si_3N_4$-$SiO_2$ (henceforth referred as $SiO_2$ for brevity), is often present as a parasitic element, but its influence on the electrical features of the Gr/Si junction is disregarded or underestimated. Here, we demonstrate that the thermal-generated minority carriers, accumulated at Si surface by the effect of the graphene gate of the Gr/SiO$_2$/Si capacitor (intrinsic gating effect), can diffuse to the Gr/Si junction and cause an anomalously high reverse leakage. Interestingly, we show that the same mechanism, when triggered by photogenerated carriers, can result in record responsivity with a photocurrent exceeding the forward current of the diode.

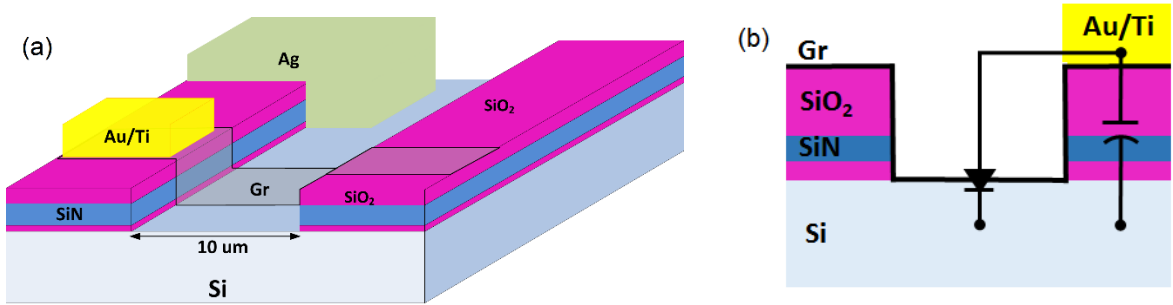

**Figure 1** (a) 3D schematic view and (b) cross-section of the device consisting of a Gr/Si diode in parallel with a Gr/SiO$_2$-Si$_3$N$_4$-SiO$_2$/Si MOS capacitor.

Furthermore, using I-V and capacitance-voltage (C-V) measurements at different temperatures, we extract relevant parameters of the device, such as barrier height, ideality factor, and series or shunt resistance.

Our study unveils an important mechanism, often neglected, which significantly contributes to the electrical transport in the Gr/S junctions. From the application viewpoint, the Gr/Si Schottky hybrid device that we propose is an excellent photodetector, with easy and cost-effective fabrication and photo-response exceeding that of semiconductor devices presently on the market.

**2. Results and discussion**

**2.1 Results**

Figure 2 (a) shows the semi-logarithmic plot of the I-V characteristic of the device, measured in dark after 2 days of vacuum annealing at $10^{-3}$ mbar. The device exhibits a clear rectifying behavior, with rectification ratio up to $10^3$, quasi-saturated reverse current and forward current exponentially growing till the reach of a downward bending region.

The shape of the I-V curve suggests using the usual diode equation for the electrical transport [4]:

$$I = I_0 \left[\exp\left(\frac{qV}{nkT}\right) - 1\right], \quad (1)$$

where n is the ideality factor, q is the electron charge, $I_0$ is the reverse saturation current (leakage), T the temperature and k the Boltzmann constant. The ideality factor accounts for mechanisms other than pure thermionic injection as well as for an unwanted insulating layer or defects inadvertently introduced at the Gr/Si interface during the fabrication process, which introduce Schottky barrier inhomogeneity. The solid blue line in Figure 2 (a) represents the I-V curve predicted by Equation (1) with the fitting parameters, $I_0$ and n, obtained from the exponential forward region $0.3 \text{ V} < V < 1 \text{ V}$ (see Electronic Supplementary Material). In such region $V \gg nkT/q$ and $V < R_s I$, where $R_s$ is the series resistance (see the following), and Equation (1) reduces to $I = I_0 \exp\left(\frac{qV}{nkT}\right)$, enabling a simple linear fitting of the semilogarithmic I-V curve to estimate $I_0 = 1.21 \times 10^{-12}$ A and $n = \frac{q}{kT} \frac{dV}{d\ln I} = 4.05$, respectively. According to the thermionic theory,

$$I_0 = AA^*T^2 \exp\left(-\frac{\Phi_{b0}}{kT}\right) \quad (2)$$

where $A = 4 \times 10^{-4} \text{ cm}^2$ is the area of the Gr/Si junction, $A^* = 112 \text{ Acm}^{-2}\text{K}^{-2}$ is the effective Richardson constant for n-Si and $\Phi_{b0}$ is the Schottky barrier height. Using the estimated value of $I_0$, Equation (2) yields

$$\Phi_{b0} = kT \ln \frac{AA^*T^2}{I_0} = 0.9 \text{ eV} \quad (3)$$

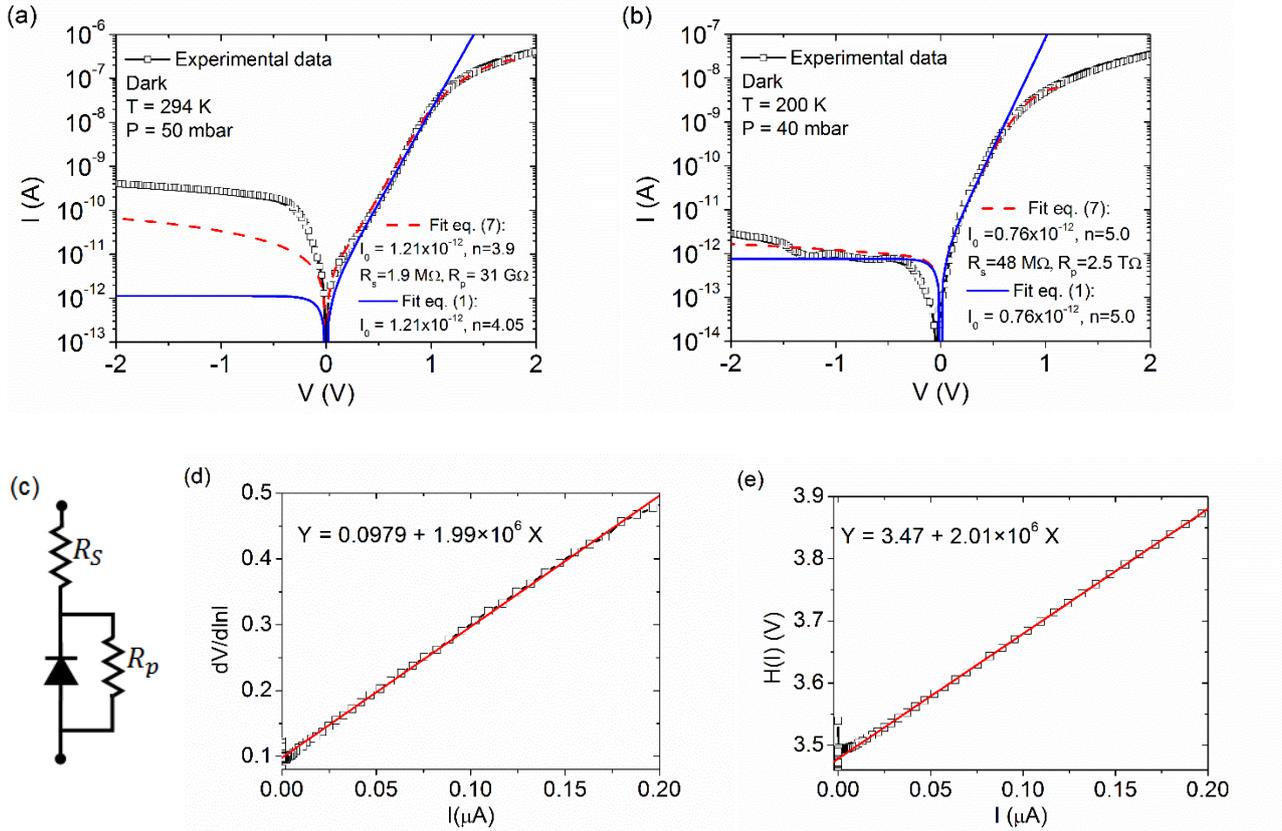

**Figure 2** I-V characteristic of the Gr/Si diode and fitting of Equation (1) (blue solid line) and Equation (7) (red dashed line) at T= 294 K (a) and T= 200 K (b), respectively. (c) Circuit schematic of the Gr/Si heterojunction with series resistance $R_s$ and shunt resistance $R_p$. (d) $dV/d\ln I$ vs I (e) and H(I) vs. V plots in the downward curvature region of the $\ln I - V$ characteristic to extract the diode parameters according to the Cheung's method [26].

The solid blue line in Figure 2 (a) shows that Equation (1) dramatically fails in reproducing the reverse saturation current, which results more than two orders of magnitude higher than the value estimated by the extrapolation of the forward current to zero bias. Indeed, a deviation from Equation (1) can be observed also at low and high forward bias, where the measured current is higher and lower than the theoretical one, respectively.

To account for these discrepancies, we consider more realistic model including a series resistance $R_s$, which is the lump sum of the resistance of substrate, graphene and contacts, and a shunt resistance $R_p$ to include possible leakages through surface, edges or defect sites (Figure 2 (c)). To estimate the series resistance $R_s$, we use Cheung's method [26]. With non-negligible series resistance, Equation (1) becomes

$$I = I_0 \left[\exp\left(\frac{q(V-R_s I)}{nkT}\right) - 1\right], \quad (4)$$

which, in the linear and downward curvature part of the forward semilogarithmic I-V curve, where $V - R_s I \gg nkT/q$, provides

$$\frac{dV}{d\ln I} = R_s I + n\frac{kT}{q} \quad (5)$$

Equation (5) is used to evaluate $R_s$ and n from the slope and the intercept of the $\frac{dV}{d\ln I}$ vs. I plot, respectively. Furthermore, defining a function H(I) as

$$H(I) \equiv V - n\left(\frac{kT}{q}\right)\ln\left(\frac{I}{AA^*T^2}\right) = R_s I + n\Phi_{b0}, \tag{6}$$

with n extracted from Equation (5), the H vs. I plot similarly yields $R_s$ (which can be used as consistency check) and $\Phi_{b0}$. This is shown in Figure 2 (d) and (e), from which the following parameters are extracted: n = 3.86, $R_s \approx 2.0$ M$\Omega$ (consistently from both equations) and $\Phi_{b0}$ = 0.90 eV. Both the ideality factor and the Schottky barrier height are in agreement with the previous estimation. Compatible values of n and $\Phi_{b0}$ are also estimated using a different approach proposed by Norde and extended by Lien [27,28], which makes use of the entire forward I-V curve, rather than relying on a limited part of it (see Electronic Supplementary Material).

When a shunt resistance $R_p$ is included, as in the circuital model of Figure 2 (c), the I-V relationship of Equation (4) is modified as

$$I = \frac{R_p}{R_s + R_p}\left\{I_0\left[\exp\left(\frac{q(V - R_s I)}{nkT}\right) - 1\right] - \frac{V}{R_p}\right\} \tag{7}$$

Equation (7), with $I_0$, $R_s$, and n previously estimated and $R_p \approx 31$ G$\Omega$ as further fitting parameter, provides an excellent fit of the forward current, as shown by the red-dashed curve of Figure 2 (a), till the reach of the flat band condition (V $\approx R_s I$) where the diode equation does not apply any more. Despite that, Equation (7) still underestimates the reverse saturation current by almost an order of magnitude (a lower value of $R_p$ could increase the reverse current but would then fail in reproducing the forward bias current). Remarkably, the same fitting procedure, applied at T = 200 K, results in a better agreement with measurements, both in forward and reverse bias (apart a slight overestimation of the reverse current close to zero bias), as displayed in Figure 2 (b).

These observations are a hint that a temperature-dependent extra current adds to the Gr/Si diode one in reverse bias.

Returning to the other diode parameters, we observe that the obtained zero bias Schottky barrier height $\Phi_{b0}$ is consistent or slightly exceeding (by ~0.1 − 0.3 eV) what has been reported for graphene on lightly doped n-Si [25,29-33]. Our estimations are based on the assumption of the theoretical $A^*$. Indeed, according to the experimental evidence, this value could be several orders of magnitude higher than the real one [16,32,34]. Although the origin of a lower effective Richardson constant is still under debate [35,36], the inadvertent presence of a native oxide layer at Gr/Si interface could explain its measured value. Indeed, an insulating layer at interface modifies Equation (2) as [37,38],

$$I_0 = AA^*T^2\exp(-\chi^{1/2}\delta)\exp\left(-\frac{\Phi_{b0}}{kT}\right) \tag{8}$$

with the introduction of a tunneling attenuation factor, $\exp(-\chi^{1/2}\delta)$, where $\chi[eV]$ is the mean barrier height and $\delta[\text{Å}]$ is the thickness of the insulating layer (the dimensional constant of $[2(2m^*)\hbar^2]^{1/2} \approx 1.01\ eV^{1/2}\text{Å}^{-1}$ is commonly omitted). The tunneling attenuation factor can be considered as a modification of the effective Richardson constant, $A^{**} = A^*\exp(-\chi^{1/2}\delta)$, and reduces the estimated $\Phi_{b0}$ in Equation (3) by an amount $-kT\,\chi^{1/2}\delta$ ( which is $\approx 0.22 - 0.45$ eV for a SiO$_2$ interfacial layer of 5-10 Å with $\chi \approx 3$ eV).

As a crosscheck, we extracted the Schottky barrier height and the effective Richardson constant from the I-V characteristics measured at different temperature (Figure 3 (a)). According to Equation (8), the zero bias saturation current is such that

$$\ln\left(\frac{I_0}{T^2}\right) = \ln\left(AA^*\exp(-\chi^{1/2}\delta)\right) - \frac{\Phi_{b0}}{kT} = \ln(AA^{**}) - \frac{\Phi_{b0}}{kT}, \tag{9}$$

and $\Phi_{b0}$ and $A^{**}$ can be obtained from the semi-logarithmic plot of $I_0/T^2$ vs. $1/T$ (Richardson plot), which is shown in Figure 3 (b). The resulting $A^{**}= 3.95\times 10^{-5}$ Acm$^{-2}$K$^{-2}$ is significantly lower than the theoretical 112 Acm$^{-2}$K$^{-2}$, and $\Phi_{b0} = 0.52$ eV confirms the overestimation of the barrier height. We remark that the previous methods would produce the same $\Phi_{b0} \approx 0.52$ eV , if we assume $A^{**}= 3.95\times 10^{-5}$ Acm$^{-2}$K$^{-2}$ and that such $A^{**}$ implies the presence of an inadvertent SiO$_2$ layer of ~8 Å at the Gr/Si interface. Indeed, X-ray photoelectron spectroscopy measurements (XPS) confirmed a SiO$_x$ layer underneath graphene slightly different from the native oxide on uncovered Si, with x<2 and thickness ~1nm (see Electronic Supplementary Material). A thin interfacial layer reduces the majority-carrier thermionic emission current without affecting the minority carrier current, which is from diffusion, thus augmenting the minority carrier injection efficiency [38]. Furthermore, the tunneling interface layer sustains part of the junction bias and introduces a voltage dependence in the Schottky barrier height, which is reflected in the appearance of the ideality factor $n = \left(1 - \frac{\partial \Phi_{b0}}{\partial V}\right)^{-1}$ different from one [37,39,40]. A barrier height of 0.52 eV is still enough to provide significant rectification, as shown in Figure 3 (c).

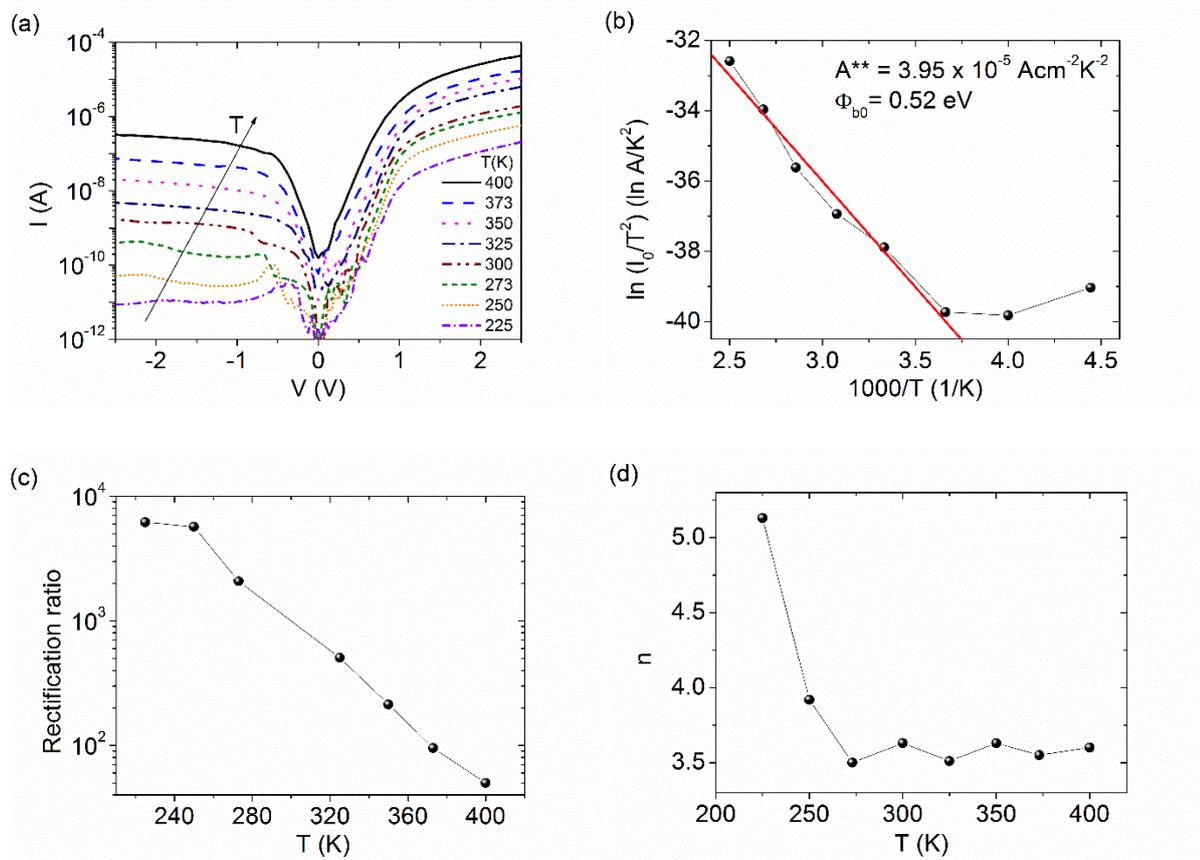

**Figure 3** (a) I-V characteristics of the Gr/Si junction for decreasing temperature from 400K to 200 K at pressure of ∼50 mbar. (b) Richardson plot to extract barrier height. (c) Rectification ratio at ∓1.5 V and (d) ideality factor n vs temperature.

The rectification ratio, here defined as the ratio of the forward to the reverse current at ∓1.5 V, decreases with raising temperature due to a faster increase of the reverse current with temperature.

Finally, Figure 3 (d) displays the temperature behavior of the ideality factor, which is ∼3.6 at room temperature and above, while increases at lower temperatures for the expected suppression of thermionic emission. The high value of n is obviously attributed to the interfacial $SiO_2$ layer. The slight overestimation of n from the fit of the linear region or from Cheung's method can be attributed to a further bias dependence of the Schottky barrier height in the forward region introduced by graphene [16].

Also the properties of the junction are strongly affected by unwanted contaminants or defects due to the fabrication process, which acts as charge traps and add interface states. These defects and contaminants, which cause charge storage and induction of charge puddles in graphene, result in local variation of the Schottky barrier height. The spatial inhomogeneity of the barrier contribute to the high n [40]. Charge trapping is suggested also by the suppression of current observed at low reverse bias at $T = 200$ K (Figure 2 (b)), where minority-carriers can be spent to fill trap states. Charge trapping is confirmed by C-V measurements as we will show in the following.

To further investigate the underlying physical mechanisms of charge transport, we tested the optical response of the device. Figure 4 (a) and (b) show the I-V characteristics at two different temperatures, under exposure to the light from a white LED system with controllable intensity. Both at room temperature and $T = 200$ K, the forward current is unchanged, while the reverse current increases dramatically, till exceeding the forward current, with increasing illumination intensity. The increase of the reverse current is particularly evident at $T = 200$ K, where the reverse current attains values almost two orders of magnitude higher than the forward ones at $5.5 \text{ mWcm}^{-2}$ illumination. This feature is quite unusual in Gr/Si photodetectors, and has not been reported or discussed before. Also, a photovoltaic effect with $V_{oc} \approx 0.2$ eV at room temperature is observed. The photocurrent at given illumination is perfectly reproducible and reaches a plateau at high reverse bias, which is the indication of limited photo-carrier generation rate. The plateau current exhibits a linear behavior with the optical power incident on the Gr/Si junction area, $P_{opt}$, as shown in Figure 4 (c). Figure 4 (d) displays the responsivity, defined as the ratio of the photocurrent to the incident power on the junction area, $\mathcal{R} = \frac{I_{ph}}{P_{opt}}$, as a function of $P_{opt}$ and shows that a peak of 3 AW$^{-1}$ is achieved at T=300 K, while is ~1 AW$^{-1}$ at T = 200 K.

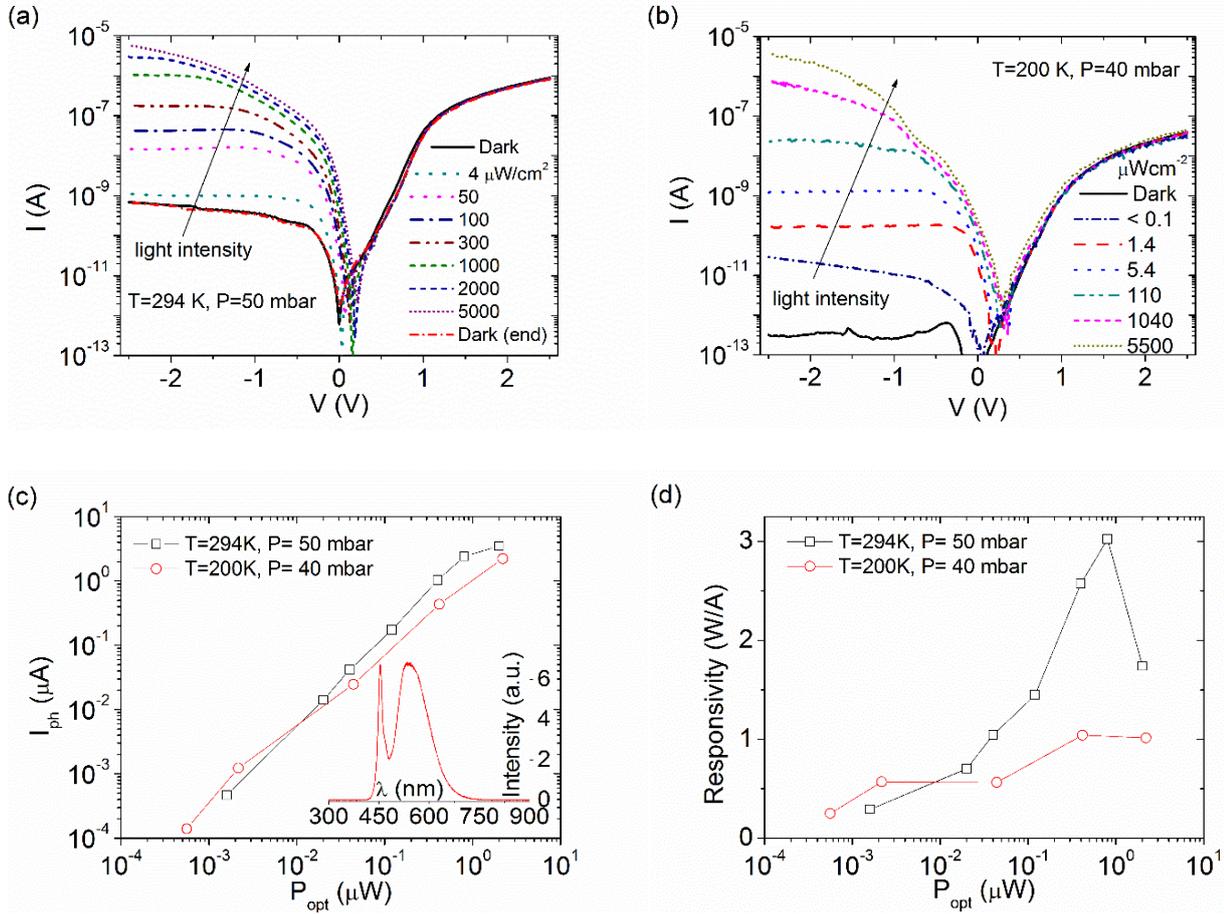

**Figure 4** I-V characteristics of the Gr/Si junction for different illumination levels at T=294 K, P=50 mbar (a) and at T = 200 K, P = 40 mbar (b) after 5 days vacuum anneal. Photocurrent (c) and responsivity (d) as a function of the optical power incident on the junction area.

The spectrum of the LED light extends over the range 420-720 nm (inset of Figure 4 (c)). Considering the average wavelength $\bar{\lambda} = 545$ nm, and assuming a constant responsivity $\mathcal{R}$ on such range, from

$$\mathcal{R} = \frac{I_{ph}}{P_{opt}} = \eta \frac{\lambda(\mu m)}{1.24} \quad [AW^{-1}], \tag{10}$$

we estimate an external quantum efficiency, which is the number of carriers produced per incident photon, $\eta > 230\%$. Such high value is in agreement with other reports on similar devices [41]. The noise equivalent power (NEP), which represents the minimum detectable optical power, defined as the rms optical power required to produce a signal-to-noise ratio of 1 in a 1 Hz bandwidth, is less than 0.1 Wcm$^{-2}$. Since the signal increases linearly with the area, while the noise varies as the square root of the area of the photodiode [38], the detectivity

$$D^* = \frac{\sqrt{AB}}{NEP} \quad [cm\ Hz^{1/2}W^{-1}] \quad (11)$$

normalized with respect to unit area (A) and unit bandwidth (B), is the commonly used figure of merit to compare different devices. For the Gr/Si photodiode under study, it results $D^* > 3.5 \times 10^{12}$ cm Hz$^{1/2}$W$^{-1}$.

We highlight that the responsivity and detectivity of our device compares with or exceeds the ones reported for similar or more complex state-of-the art graphene based photodetectors [16,42-46].

**2.2 Discussion**

To understand the observed electrical and opto-electrical features of the device, we have to carefully consider the role of the parallel Gr/SiO$_2$/Si MOS capacitor (Figure 5 (a) and (b)). Sideways to the Gr/Si junction region, graphene acts as the gate of the MOS structure and causes accumulation, depletion or inversion of the underneath Si substrate. We call this effect internal gating of the device. The Gr/Si junction and the MOS capacitor feel the same applied voltage. In forward bias (Figure 5 (c)), the positive voltage on graphene causes accumulation of electrons at the Si/SiO$_2$ interface of the MOS capacitor (charge can be stored also in SiO$_2$ traps). Such electrons do not affect much the I-V characteristic of the device since the forward current of the Gr/Si diode is limited by thermionic injection rate over the barrier rather than by the availability of free carriers and adding extra electrons does not make a difference. On the contrary, in reverse bias (Figure 5 (d)), when free electrons cannot overcome the barrier, the reverse saturation current is dominated by minority carriers (holes) and is strongly controlled by their availability, i.e. generation rate. When the reverse-biased graphene gate attracts thermally- or photo-generated holes at Si/SiO$_2$ interface, which becomes a minority carrier reservoir, the Gr/Si junction is provided with an extra source of minority carriers (Figure 5 (d)).

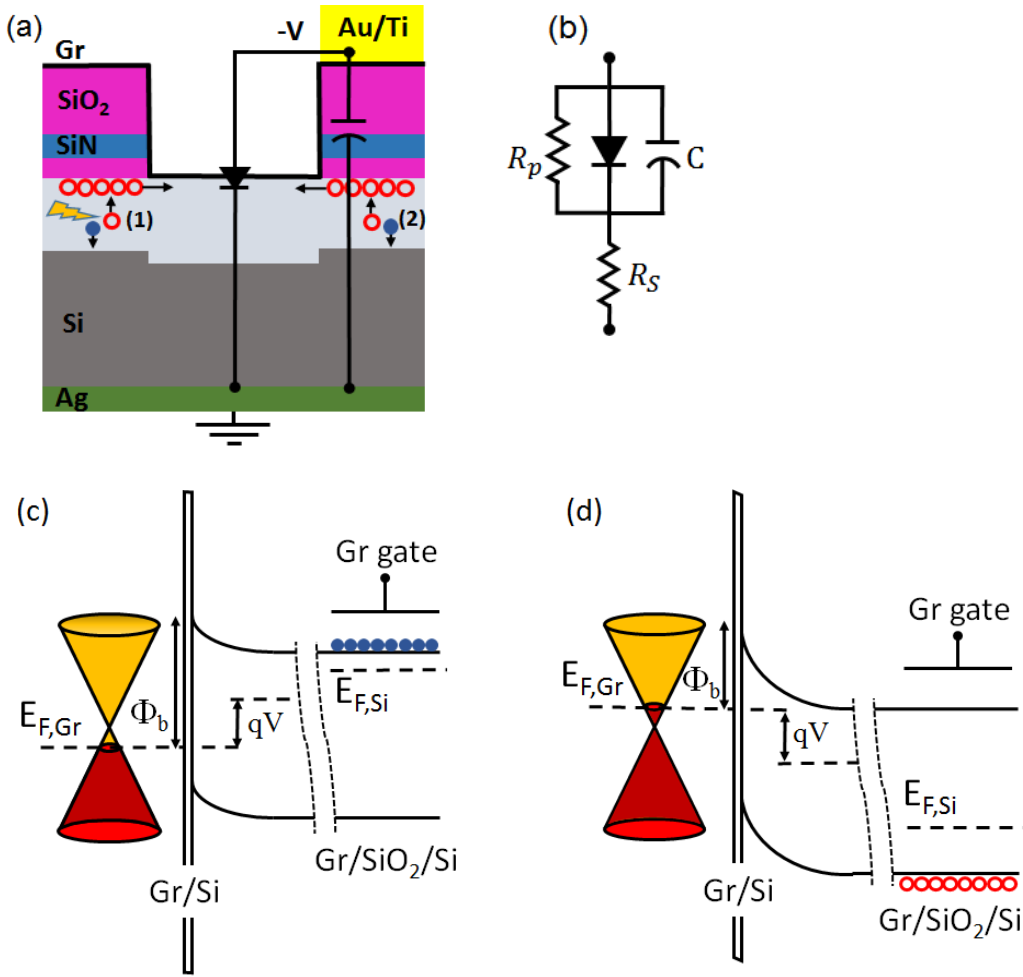

**Figure 5** (a) Layout of the device under study, consisting of a Gr/n-Si junction in parallel with a Gr/SiO$_2$-SiN-SiO$_2$/n-Si MOS capacitor. Photo- (1) and thermo- (2) generation of charge carriers are illustrated. (b) Circuit equivalent of the device. (c) and (d) Band diagram in the Gr/n-Si junction and in the Gr/SiO$_2$ /n-Si MOS capacitor region, in forward and reverse bias, respectively.

Indeed, such carriers diffusing to the junction region are suddenly swept by the high electric field of the reverse biasing and contribute to the reverse current. Moreover, the ultrathin SiO$_2$ layer at the Gr/Si junction is transparent to them, as already pointed out. It is also important to realize that the biasing of the Gr/Si junction reduces the barrier against the diffusion of the minority carriers accumulated at the MOS region, which are then favored to move to the junction area. This is at the origin of the unusually high reverse saturation current observed at room temperature in dark and of the extraordinarily high photocurrent. At lower temperature, the thermal generation is suppressed and so is the rate of diffusion from the MOS region into the junction area. This makes the junction reverse current much closer to the theoretical value and the photocurrent slightly lower at T = 200 K (Figure 4 (c)).

To experimentally validate the model, we carried out C-V characterization in dark and under illumination. We performed small-signal capacitance measurements as a function of the bias and at different frequencies, both on the device under study and on a test structure, on the same chip, consisting of a small-area MOS capacitor with Ag top electrode and $SiO_2$-$Si_3N_4$-$SiO_2$ dielectric (referred as Ag/$SiO_2$/Si MOS for brevity). As displayed in Figure 6 (a), the capacitance of the Ag/$SiO_2$/Si test structure at the high frequency of 10 kHz exhibits the usual MOS transition from accumulation with high capacitance in forward bias, to depletion in the range 1 to 0 V with decreasing capacitance, to inversion with lower capacitance at V < 0 V. In reverse bias, the minority carriers of the inversion layer are not able to follow the fast 10 kHz oscillations of $V_{ac}$ and the charge of the gate is mirrored by variations of the depletion layer thickness. Hence, the value of the capacitance of the insulating layer, measured in accumulation, is decreased by the depletion region capacitance which adds to it in series. Under illumination and at the lower frequency of 1 kHz, the inversion layer at the Si/$SiO_2$ interface is able to respond to the variation of the ac signal, $V_{ac}$, and the capacitance recovers the high value observed in accumulation, as shown by the red curve in Figure 6 (a). The photo-generation rate is able to provide the positive charge needed in the inversion layer to follow the oscillations of the small ac signal.

The bias dependence of the small-signal capacitance of the Ag/$SiO_2$/Si MOS test structure confirms the hypothesis that a p-type inversion layer is formed upon illumination (or for thermal effect) at the $SiO_2$/Si interface of the Gr/$SiO_2$/Si of the device under study, over the used bias range. As explained, such inversion layer contributes to the high reverse current and photocurrent. We highlight that the relatively high flat-band potential $V_{FB} \approx 1V$ observed on the C−V curve of the MOS test structure indicates the presence of trap states filled with negative charge.

Figure 6 (b) displays the $C-V$ characteristics of the device under test. The curve in dark, for $V < 1V$, has similar features to the capacitance of the Ag/SiO$_2$/Si MOS test structure, with a transition from higher (accumulation) to lower (inversion) capacitance. Such observation suggests that the capacitance of the parallel Gr/SiO$_2$/Si MOS dominates over the Gr/Si junction capacitance, due to its larger area. However, a closer look at the $C-V$ curve in dark of Figure 6 (b) shows a local minimum at V~0 V, which should be the inversion point, and that the lower value of the capacitance is attained at $V < -2V$. A shoulder appears in the range $-2V < V < 0$ V, where it is likely that the growing reverse-biased Gr/Si junction capacitance overcomes the Gr/SiO$_2$/Si MOS capacitance. This interpretation is confirmed by the linear behavior of the $1/C^2 - V$ plot shown in Figure 6 (c). In fact, for a Schottky non-ideal diode the inverse square of reverse-bias capacitance is a linear function of the bias [39,47]:

$$\frac{1}{C^2} = \frac{2\,n\,[n\,(\Phi_{b0} - \Phi_n - kT) - qV]}{A^2 q^2 \varepsilon_s N_d} \tag{12}$$

Where $N_d$ is the doping density, $\varepsilon_s = 11.7\varepsilon_0$ is the silicon permittivity, and $\Phi_n = kT\ln\frac{N_c}{N_d}$ with $N_c = 12\,(2\pi m_n kTh^{-2})^{3/2}$ the effective density of states in the conduction band ($= 2.9 \times 10^{19}$ cm$^{-3}$ at T = 300K; $m_n = 0.25 m_e$ is the effective electron mass, h is the Planck constant). According to Equation (12), the x-intercept, $V_0$, of the straight-line fitting the $1/C^2 - V$ curve can be used to evaluate the barriers height, $\Phi_{b0}$, as

$$\Phi_{b0} = \frac{V_0}{n} + kT\ln\left(\frac{N_c}{N_d}\right) + kT \approx 0.55 \text{ eV} \tag{13}$$

with $N_d = 4.5 \times 10^{14}$ cm$^{-3}$ and n = 3.6.

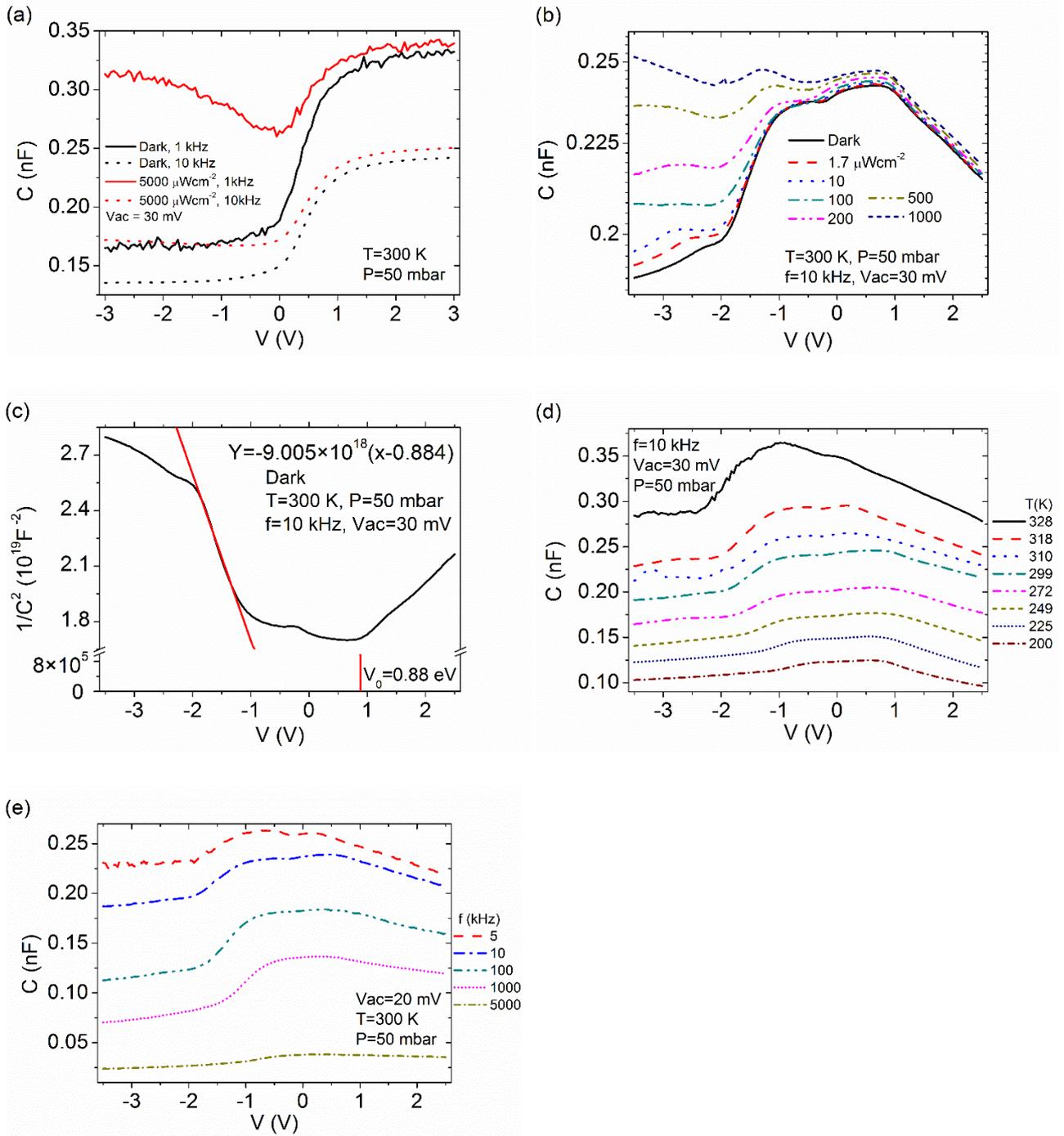

**Figure 6** Small-signal capacitance-voltage, $C-V$, measurements at 1 and 10 kHz in dark and under illumination for (a) a Ag/SiO$_2$/Si MOS test structure and (b) for the hybrid device under study made of the parallel of Gr/Si junction and Gr/SiO$_2$/Si MOS capacitor. (c) $1/C^2 - V$ plot for the hybrid device. (d) Temperature dependence of C – V curves and (e) frequency dependence of the C – V curves for the hybrid device.

The value of $\Phi_{b0}$ so obtained is in good agreement with the barrier height extracted with Richardson's method using the data of Figure 3. Equation (12) also shows that the slope of the $1/C^2 - V$ curve can be used to evaluate the doping density $N_d$:

$$N_d = -\frac{2nA^{-2}}{\varepsilon_s q \frac{d(1/C^2)}{dV}} \approx 3\times 10^{19} \text{ cm}^{-3} \qquad (14)$$

The resulting value is 5 orders of magnitude higher than the nominal $N_d$, confirming that in reverse bias the Si substrate goes in inversion ($N_d$ extracted from Equation (14) is indeed the hole density in the inversion layer).

The increase of the reverse bias-capacitance with light intensity shown in Figure 6 (b) is likely due to the restoring of the ability of the inversion layer of the MOS capacitor to follow the high frequency oscillations, favored by the high electron-hole pair photogeneration rate. In forward bias, for $V >$ 1V, the capacitance drop is due to the series resistance $R_s$ and the current I through the forward biased Gr/Si diode [48,49,50]. Interface trap states are expected at the Si/SiO$_2$ interface, either in the junction or MOS region of the device. The non-stoichiometric SiO$_x$ layer evidenced by XPS analysis indicates that a larger density of interface trap states is located at Gr/Si junction. The capacitance due to charge storage in the interface trap states ($C_{it}$) is in parallel with the depletion layer capacitance ($C_{sc}$) of the Gr/Si junction, therefore adds to it. Both $C_{it}$ and $C_{sc}$ are in series with the SiO$_x$ layer capacitance ($C_i$), which in turn can include trap states. At high frequencies, the forward capacitance of a Schottky diode can be expressed as [51]:

$$C = \frac{C_{sc}C_i}{C_i + C_{sc} + R_s C_i qI/kT}, \qquad (15)$$

and decreases with increasing current, giving rise to the sort of square-pulse shaped peak observed in Figure 6 (b) in dark. We notice that a terms $R_s C_i qI/kT$ appears also in the denominators of the expression of C at low frequency [51].

Interface trap states of the diode and MOS regions are expected to introduce a temperature and frequency dependence in the capacitance of the device, since temperature and frequency strongly affect their dynamic [47,52,53]. This is confirmed by Figure 6 (d) and (e), which show the temperature and frequency dependence of the measured capacitance.

The dependence of C on temperature is an indication of the presence of a continuous distribution of the interface states in the bandgap of the semiconductor. Charge trapping is an energy activated process and strongly depends on temperature. Lowering the temperature suppresses the probability of trap filling, thus decreasing the traps contribution to the capacitance [54,55]. Similarly, due to relaxation mechanisms at the interface, the capacitance is frequency-dependent. While at lower frequencies the interface states follow the frequency, at higher frequencies they are not able to follow the potential oscillations and the interface states do not contribute to the capacitance [52,56,57]. This occurs when the time constant is too long to permit the charge to move in and out of the trap states in response to the applied ac signal. Accordingly, the interface states capacitance has been neglected in Figure 6 (a)-(c) since measurements are done at high frequency (10 kHz) at which traps should poorly follow the applied ac signal.

## 3. Conclusion

In summary, we have proposed a hybrid device consisting of a Gr/Si junction in parallel with a Gr/SiO$_2$/Si MOS capacitor, which demonstrates itself as a high responsivity/detectivity photodetector. The device is easy to fabricate and to integrate into the existing semiconductor technologies for graphene-based optoelectronic applications.

We have performed a detailed I-V and C-V characterization at different temperatures to extract relevant parameters of the Schottky diode and investigate the underlying physics. We have shown the effect of the often-neglected Gr/SiO$_2$/Si parasitic capacitor on the I-V characteristics of the Gr/Si junction. We have unveiled the key role of photo-generated minority carries, accumulated in the capacitor region, to enhance the photoresponse of the device to a level of endowing it with a photocurrent exceeding the forward current.

## 4. Experimental

Samples were prepared on doped n-Si (100) wafers with the resistivity of about 10 Ωcm corresponding to a phosphorus dopant density of ∼4.5×$10^{14}$ cm$^{-3}$. The fabrication flow is illustrated in Figure 7.

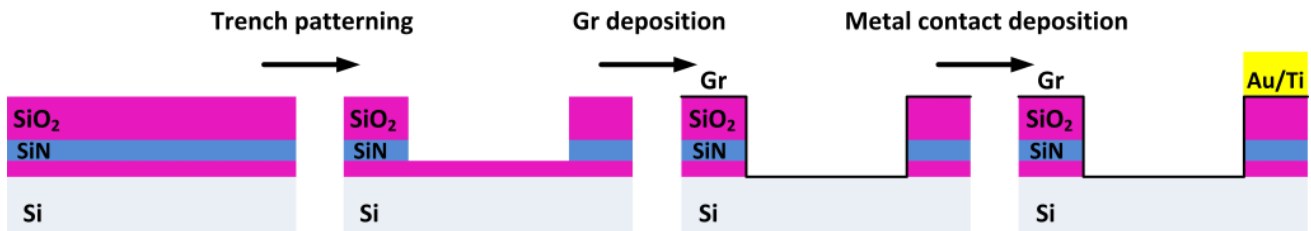

**Figure 7** Sample fabrication flow.

Preparation starts with the formation of a 15 nm-thick thermal SiO$_2$ layer which is followed by the chemical vapor deposition (CVD) of 30 nm-thick silicon nitride and a 200 nm SiO$_2$ layer. In the next step, a 10 μm-wide trench is patterned by lithography, dry SiO$_2$ etching and silicon nitride wet etch. The 15 nm-thick thermal SiO$_2$ is removed from the trench area with a hydrofluoric acid dip immediately before the graphene (Gr) deposition. This reduces the time exposure of bare Si surface to air, thus preventing or limiting the formation of native oxide. The Gr/Si junction and the parallel Gr/SiO$_2$/Si capacitor are formed simultaneously by transferring ∼1 × 0.4 cm² CVD-graphene sheets from Cu foils onto the Si substrates, following the wet etch method described in detail elsewhere [58]. Graphene is placed to cover the Si trench and the surrounding oxide stack (which, after etching, is ∼150 nm thick), thus acting both as anode of Gr/Si junction and top (gate) electrode of the Gr/SiO$_2$/Si MOS capacitor. Ohmic metal contact to graphene is fabricated by evaporation of Ti/Au metal stack through a shadow mask [59]. Finally a top-contact to the Si-substrate is established by depositing Ag paste on the exposed areas of the wafer, opportunely scratched to guarantee ohmic contact.

Figure 8 (a) shows a scanning electron microscopy (SEM) image of the trench edge after Gr transfer. The characteristic graphene wrinkles extending across the edge of the trench from the isolated area as well as on the Gr/Si junction area are visible. Gr layer covers the edge of the trench keeping its structural integrity and ensuring electrical contact between the junction area and the contact area on top of the isolation layer. A darker region (~ 100 ÷ 200 nm wide) extending along the trench wall indicates that graphene may be free-standing in this region. As the latter area constitutes a negligible fraction of the Gr/Si junction area it is assumed for simplicity that the entire trench region is uniformly covered with Gr. Figure 8 (b) shows an optical microscope image of the trench covered with Gr. The optical contrast is significantly worse than that on $SiO_2$ and Gr layer cannot be clearly identified [58]. In contrast, a Raman line-scan (Figure 8 (c)) performed across the trench provides a clear evidence of a high quality Gr layer. The D-peak (~1350 cm$^{-1}$) is negligible both on the isolation and junction areas. The significantly better signal-noise ratio for spectra acquired on the $SiO_2$ isolation is a consequence of the interference enhancement of Raman Gr signals over silica [60,61].

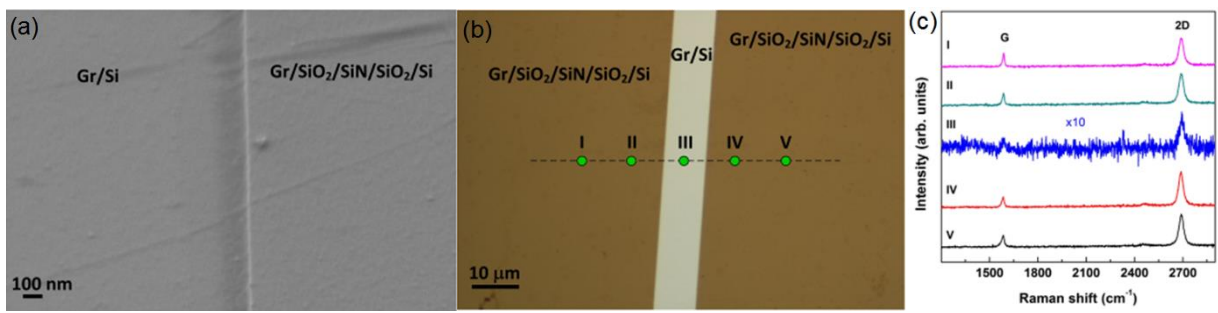

**Figure 8** (a) SEM image of the trench edge covered with graphene. (b) Optical microscope image of the trench. The junction area is $4 \times 10^{-4}$ cm$^2$, while the total area of the of the device under study is ~0.4 cm$^2$. (c) Raman spectra from a line scan across the trench as indicated in panel (b). Signal for region III is amplified (×10) for clarity.

Electrical and optical measurements were performed in a high-vacuum, cryogenic Janis probe station connected to a Keithley 4200 SCS parameter analyzer. The forcing bias was applied to the Au/Ti contact on graphene, while the Ag pad on Si was grounded. To avoid moisture on the surface, the sample was subjected to a vacuum annealing (<10$^{-3}$ mbar) for two days or more, and kept below 50 mbar in rest gas atmosphere during the measurements. For the capacitance measurement, we utilized the parallel model for which results are expressed as the parallel capacitance (C) and the parallel conductance (G).

**Electronic Supplementary Material**: Supplementary material (diode parameter extraction, Cheung's method to evaluate $R_s$, n and $\Phi_{b0}$ at T=200 K, modified Norde's method to evaluate $R_s$, n and $\Phi_{b0}$ at T=300 K, further C-V measurements, and X-ray photoelectron spectroscopy on the junction region) is available in the online version of this article at http://dx.doi.org/10.1007/**********************).